\documentstyle[12pt,epsfig]{article}
\def\lapprox{\lower .7ex\hbox{$\;\stackrel{\textstyle <}{\sim}\;$}}
\def\gapprox{\lower .7ex\hbox{$\;\stackrel{\textstyle >}{\sim}\;$}}

\def\d{{\rm d}}

\def\barq{\bar{q}}

\def\xa{x_{1}}
\def\xb{x_{2}}
\def\xaa{x_{1}^{0}}
\def\xbb{x_{2}^{0}}

\def\Li{{\rm Li}}

\begin{document}
\begin{titlepage}
\vspace*{-1cm}
\begin{flushright}
DESY--97--205   \\
October 1997 \\
\end{flushright}                                
\vskip 3.5cm
\begin{center}
{\Large\bf QCD Corrections to Double and Single}
\vskip .1cm
{\Large\bf Spin Asymmetries in Vector Boson Production at}
\vskip .1cm
{\Large\bf Polarized Hadron Colliders}
\vskip 1.cm
{\large  T.~Gehrmann} 
\vskip .4cm
{\it DESY, Theory Group, D-22603 Hamburg, Germany}
\end{center}
\vskip 3.cm

\begin{abstract}
The production of $W^{\pm}$ and $Z^0$ 
bosons in collisions of longitudinally polarized 
protons 
allows to access various combinations of the polarized 
quark distributions from the study of double and single spin asymmetries.
We compute the ${\cal O}(\alpha_s)$ corrections to these asymmetries 
as function of the vector boson rapidity $y$ in the hadron--hadron 
centre-of-mass frame.
Detailed studies of the numerical 
impact of the next-to-leading order
corrections are presented. The theoretical uncertainties of an extraction 
of polarized quark distributions from future vector boson production data are 
investigated in detail. 
\end{abstract}


\vfill
\end{titlepage}                                                                
\newpage                                                                       

\section{Introduction}

The experimental
knowledge on the spin structure of the nucleon has up to now 
been largely restricted to measurements~\cite{g1exp} 
of the polarized structure function 
$g_1^{p,n,d} (x,Q^2)$. This structure function probes a particular 
charge-weighted combination of the polarized quark distributions. Its 
mere knowledge is therefore insufficient to disentangle the contributions 
of valence and sea quarks to the nucleon spin and for a further 
decomposition of the light quark sea into different flavours.  
A recent study of asymmetries in semi-inclusive hadron production 
by the SMC collaboration~\cite{smchadr}
yielded some additional constraints on the 
valence and sea quark polarization. These data 
are however still insufficient for a simultaneous determination 
of all different polarized quark and antiquark distributions. 

One of the major future projects in the study of the  
nucleon spin structure is the operation of the Relativistic 
Heavy Ion Collider (RHIC) at BNL with 
two colliding polarized proton beams at centre-of-mass energies of
$\sqrt S = 200 \ldots 500$~GeV~\cite{rhic}. The RHIC spin programme 
is expected to start operation three years from now and will offer 
new information on the polarized parton distributions from a 
variety of observables accessible in hadron--hadron 
collisions~\cite{sofferrep}. A recent overview of the physics prospects 
of the RHIC spin programme has been presented in~\cite{soffernew}, where 
estimates of the expected asymmetries and anticipated errors for  
the most important processes can be found.  At RHIC it will be 
possible to access the polarized quark distributions from the study of
double and single spin asymmetries in the production of 
$V=W^{\pm},Z^0$ vector bosons.

Up to very recently, these asymmetries have only been 
studied~\cite{sofferrep,soffernew,leader,soffer} at lowest order 
in perturbation theory, although it is well known from the 
unpolarized case that  perturbative 
corrections to 
vector boson production cross sections at hadron colliders 
are sizable~\cite{dym1,dym2,aem,kubar,wtrans,wdecay,wjets}. The calculation 
of the next-to-leading order (${\cal O}(\alpha_s)$)
corrections to single and double spin 
asymmetries as function of the vector boson rapidity $y$
in the hadron--hadron centre-of-mass frame
and the study of their numerical impact is the aim of the 
present paper. The perturbative
corrections to double spin asymmetries in vector boson 
production are identical to the corrections in the longitudinally 
polarized Drell--Yan process and can therefore be easily obtained 
from~\cite{weber,tgdy}. Corrections to the corresponding 
single spin asymmetries have been first considered in~\cite{weber2} in 
the context of a soft gluon resummation. We shall present 
a rederivation of these corrections, which follows closely our
earlier calculation of the QCD corrections to the double spin asymmetry 
in the Drell--Yan process~\cite{tgdy}. 

The QCD corrections to the 
total asymmetries (integrated over the vector boson rapidity $y$)
in Drell--Yan process and vector boson production at RHIC have 
recently been studied in~\cite{kamal1,kamal2}. Comparison with these 
provides a strong cross-check of our results. 

A fully consistent numerical 
study of spin asymmetries in vector boson production
 at next-to-leading
order was until now not possible, as the polarized 
parton distributions were only determined at leading accuracy. With the 
recently calculated polarized two--loop splitting functions~\cite{nlosplit},
the polarized distributions can now be determined to next-to-leading order
from fits~\cite{gs,grsv,bfr,allfits} 
to polarized structure function data. Having a 
complete calculation of the two-loop splitting functions available, it is 
now furthermore possible to define consistent scheme
transformation prescriptions~\cite{nlosplit,bfr} for parton
distributions, splitting functions and parton level cross sections at 
next-to-leading order. Possible choices of the factorization scheme 
in polarized and unpolarized Drell--Yan process are elaborated 
in great detail in~\cite{kamal1,kamal2}. In our study, we 
shall only work in the $\overline{{\rm MS}}$ scheme (with an appropriate 
correction for spurious terms generated due to the chosen representation 
of $\gamma_5$), which is identical to the scheme used in the recent 
determinations of polarized parton distributions at next-to-leading 
order~\cite{gs,grsv}.

In the context of the present study, we shall only consider the 
rapidity distributions of the produced vector bosons, not of their 
decay products. In the case of $Z^0$ boson production, 
identified from the dilepton decay mode, a reconstruction of the boson  
rapidity is indeed possible, such that a direct comparison of our results 
with experimental data is feasible. $W^{\pm}$ bosons can on the other hand
only be detected 
from their $l\nu_l$ decay mode, where just the lepton is observed, thus 
rendering a direct reconstruction of the $W^{\pm}$ rapidity unfeasible. 
The calculation presented here for the $W^{\pm}$ bosons is therefore 
only a first step towards a prediction of the lepton plus missing transverse 
momentum distribution in polarized hadronic collisions at next-to-leading 
order.

This paper is structured as follows: in section~\ref{sec:pheno}, 
we define the single and double spin asymmetries accessible in vector 
boson production at RHIC and review briefly their phenomenology. 
Section~\ref{sec:calc} contains the 
calculation of the QCD corrections to these asymmetries. 
As this calculation is 
technically 
very similar to the calculation of the corrections to the polarized 
Drell--Yan process presented in our earlier work~\cite{tgdy}, we will 
only   
point out the differences to the Drell--Yan calculation without 
presenting all details of our derivation. The numerical magnitude of the 
corrections will be studied in detail in section~\ref{sec:num}. Finally,
section~\ref{sec:conc} contains a summary of our results and concluding 
remarks.

\section{Spin Asymmetries in Vector Boson Production}
\label{sec:pheno}

The production of vector 
bosons in hadronic collisions 
can be described accurately in the narrow-width approximation
(e.g.~\cite{wjsbook}) as production of an on-shell massive vector boson by 
quark-antiquark annihilation. Like in the case of the polarized Drell--Yan 
process, one can study the double spin asymmetry 
as a function of the rapidity of the vector boson in 
the hadronic centre-of-mass frame: 
\begin{equation}
A_{LL}(y) \equiv  \frac{\d \Delta \sigma_{LL}}{\d y}\, \Bigg/ \,
\frac{\d \sigma}{\d y}, 
\label{eq:defll}
\end{equation}
where 
\begin{eqnarray}
\d \Delta \sigma_{LL} & = & \frac{1}{4} \left( \d \sigma^{++} - \d \sigma^{+-} 
- \d \sigma^{-+} + \d \sigma^{--}\right)\; ,  \label{eq:sigll} \\
\d \sigma & = & 
\frac{1}{4} \left(\d \sigma^{++} + \d \sigma^{+-} 
+ \d \sigma^{-+} + \d \sigma^{--} \right)\; , \label{eq:sig0} 
\end{eqnarray}
with $(+)$ and $(-)$ denoting positive and negative hadron helicities. 

\begin{figure}[t]
\begin{center}
~ \epsfig{file=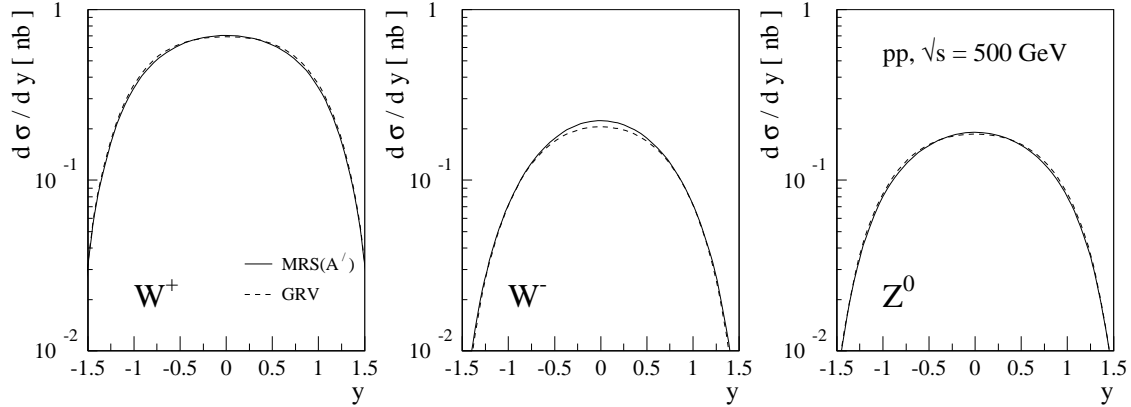,height=15cm,angle=-90}
\caption{
Unpolarized vector boson production cross sections in proton-proton 
collisions at RHIC ($\sqrt S = 500$~GeV). 
}
\label{fig:vectot}
\end{center}
\end{figure}
The unpolarized cross section in the denominator of this asymmetry 
can be predicted using the known unpolarized parton distribution functions. 
It is evaluated at next-to-leading order 
 for $pp$ collisions at $\sqrt{S}=500$~GeV in 
Fig.~\ref{fig:vectot}, using the unpolarized parton distributions 
from MRS(set A$'$)~\cite{mrsap} and GRV~\cite{grv}. The rather 
small discrepancy between both predictions seems to suggest that 
the unpolarized cross section can be predicted reliably. This is however 
not true in reality, since the ratio of $W^+$ and $W^-$ production 
cross sections in proton-proton collisions is proportional to the ratio of the 
unpolarized light antiquark distributions $\bar d(x,Q^2)/\bar u(x,Q^2)$, 
which is only poorly known at present. In principle, a measurement of 
the unpolarized $W$ boson production cross sections at RHIC could help 
to determine this ratio, as suggested in~\cite{soffer,sofferpdf}. However,
by the time RHIC comes into operation, this ratio should already
be well constrained
from measurements of the Drell--Yan lepton pair production in 
proton-proton and proton-deuterium collisions~\cite{seaasy,na51}, 
thus enabling a reliable prediction of the unpolarized cross sections.

The phenomenology of the double spin asymmetries is best understood 
at leading order (and neglecting the CKM mixing factors $|V_{ij}|$ for 
$W^{\pm}$ production), where they can be expressed~\cite{leader,soffer}
as ratios of polarized and unpolarized quark distributions
\begin{eqnarray}
A^{W^+}_{LL} (y)& = & - \frac{\Delta u(\xaa,M_W^2) \Delta \bar d(\xbb,M_W^2) +
\Delta \bar d(\xaa,M_W^2) \Delta u(\xbb,M_W^2)}{u(\xaa,M_W^2) 
\bar d(\xbb,M_W^2) + \bar d(\xaa,M_W^2) u(\xbb,M_W^2)}\; , \nonumber \\
A^{W^-}_{LL} (y)& = & - \frac{\Delta d(\xaa,M_W^2) \Delta \bar u(\xbb,M_W^2) +
\Delta \bar u(\xaa,M_W^2) \Delta d(\xbb,M_W^2)}{d(\xaa,M_W^2) 
\bar u(\xbb,M_W^2) + \bar u(\xaa,M_W^2) d(\xbb,M_W^2)}\; ,  \\
A^{Z^{0}}_{LL} (y) & = & - \frac{\sum_q (v_q^2 + a_q^2) 
\left[ \Delta q(\xaa,M_Z^2) \Delta \bar q(\xbb,M_Z^2) + 
\Delta \bar q(\xaa,M_Z^2) \Delta q(\xbb,M_Z^2) \right] }
{ \sum_q (v_q^2 + a_q^2) 
\left[  q(\xaa,M_Z^2) \bar q(\xbb,M_Z^2) + 
\bar q(\xaa,M_Z^2)  q(\xbb,M_Z^2) \right] }\; ,\nonumber
\end{eqnarray}
the distribtions being probed at 
\begin{equation}
x_{1,2}^0 = \sqrt{M_V^2/S}\, e^{\pm y}. 
\label{eq:x0def}
\end{equation}

At RHIC with $\sqrt{S}=500$~GeV, these asymmetries are thus sensitive on 
$x_{1,2} \gapprox 0.05$, where
one expects the polarized quark distributions 
$\Delta u(x,Q^2)$ and $\Delta d(x,Q^2)$ to be dominated by 
contributions from valence quarks, which are reasonably well constrained 
from inclusive and semi-inclusive asymmetries in lepton--hadron 
scattering. The above double spin 
asymmetries can therefore be used as a direct probe of the polarized 
antiquark distributions. Even the flavour structure of the polarized quark 
sea can be revealed, since $W^+$ production probes only $\Delta \bar d$, while 
$W^-$ production is particularly sensitive on $\Delta \bar u$. 

In addition to the above double spin asymmetries, it is also possible to 
define non-zero single spin asymmetries, which are induced by the 
parity violating couplings of the vector bosons. These asymmetries 
are absent in the (parity conserving) classical Drell--Yan process and 
their measurement 
requires 
only one of the incoming hadrons to be polarized. Taking 
hadron~1 (moving in the $+y$ direction) to be polarized and 
hadron~2 to be unpolarized, one can 
define the following single spin asymmetry 
\begin{equation}
A_{L}(y) \equiv  \frac{\d \Delta \sigma_{L}}{\d y}\, \Bigg/ \,
\frac{\d \sigma}{\d y}, 
\label{eq:defl}
\end{equation}
where 
\begin{equation}
\d \Delta \sigma_{L} = \frac{1}{4} \left( \d \sigma^{++} 
+ \d \sigma^{+-} - \d \sigma^{-+} - \d \sigma^{--} \right)\;  
\label{eq:sigl}
\end{equation}
and $\d \sigma$ as in eq.(\ref{eq:sig0}) above. 

At leading order (again neglecting the CKM mixing 
factors for $W^{\pm}$ production), one finds the 
following simple expressions for these single spin 
asymmetries~\cite{leader,soffer}:
\begin{eqnarray}
A^{W^+}_{L} (y)& = & \frac{-\Delta u(\xaa,M_W^2) \bar d(\xbb,M_W^2) +
\Delta \bar d(\xaa,M_W^2) u(\xbb,M_W^2)}{u(\xaa,M_W^2) 
\bar d(\xbb,M_W^2) + \bar d(\xaa,M_W^2) u(\xbb,M_W^2)}\; , \nonumber \\
A^{W^-}_{L} (y)& = & \frac{- \Delta d(\xaa,M_W^2) \bar u(\xbb,M_W^2) +
\Delta \bar u(\xaa,M_W^2) d(\xbb,M_W^2)}{d(\xaa,M_W^2) 
\bar u(\xbb,M_W^2) + \bar u(\xaa,M_W^2) d(\xbb,M_W^2)}\; , \nonumber \\
A^{Z^{0}}_{L} (y) & = & \frac{\sum_q (v_q^2 + a_q^2) 
\left[- \Delta q(\xaa,M_Z^2) \bar q(\xbb,M_Z^2) + 
\Delta \bar q(\xaa,M_Z^2) q(\xbb,M_Z^2) \right] }
{ \sum_q (v_q^2 + a_q^2) 
\left[  q(\xaa,M_Z^2) \bar q(\xbb,M_Z^2) + 
\bar q(\xaa,M_Z^2)  q(\xbb,M_Z^2) \right] }\; .
\end{eqnarray}

At large and positive rapidity $y$, ($\xaa > \xbb$), the above 
expressions are dominated by the first term in the numerator, since 
$\Delta q(x_1^0,Q^2) \gg \Delta \bar q(x_1^0,Q^2)$ for large 
$x_1^0$. The second term in the numerator
is dominant for large and negative $y$, ($\xaa < \xbb$), since 
$q (x_2^0,Q^2) \gg \bar q(x_2^0,Q^2)$ for large $x_2^0$.  
In practice, this means that a measurement of these 
single spin asymmetries at RHIC with $\sqrt{S}=500$~GeV can 
probe~\cite{soffernew,soffer} the 
polarized quark distributions for $x\gapprox 0.2$ from $A_{L} (y>0)$
and the polarized 
antiquark distributions for $x\lapprox 0.12$ from $A_L(y<0)$.
A discrimination of different 
quark flavours is again possible from a study of $W^+$ and $W^-$ 
boson production. 

Using the 
double and single spin asymmetries $A_{LL}(y)$, $A_{L}(y)$, $A_{L}(-y)$ 
together with the unpolarized cross section $\d \sigma(y)/\d y$, 
one can calculate the cross section $\d \sigma^{\lambda_1\lambda_2}
(y)/\d y$ for any combination $\lambda_1\lambda_2$
of hadron helicities. It is therefore 
possible to express the parity violating double spin asymmetries 
$A_{LL}^{{\rm PV}}(y)$ and $\overline{A}_{LL}^{{\rm PV}}(y)$ discussed 
in~\cite{soffernew} by linear combinations of the asymmetries defined 
above. 

\section{Perturbative corrections to the polarized vector boson 
production cross sections}
\label{sec:calc}

The derivation of the ${\cal O}(\alpha_s)$ corrections to spin 
asymmetries in hadronic vector boson production is in many aspects 
similar to the calculation of the ${\cal O}(\alpha_s)$ corrections 
to the $y$ distribution of lepton pairs produced via the Drell--Yan 
process in collisions of longitudinally polarized hadrons. We have 
recently presented a complete derivation~\cite{tgdy} of these 
corrections to the longitudinally 
polarized Drell--Yan process, 
which was following 
closely the unpolarized calculation of
Altarelli, Ellis and Martinelli~\cite{aem}. 

The only lowest order contribution to the Drell--Yan process 
comes from the annihilation of a quark-antiquark 
pair into a virtual photon $\gamma^*$ (or on-shell vector boson $V$)
of invariant mass $M^2$. At next-to-leading 
order, one has two different contributions: the ${\cal O}(\alpha_s)$
correction (emission of a real or virtual gluon) to the $q\bar q$
annihilation process and the quark-gluon compton scattering 
process $qg\to q \gamma^*(V)$, where $q$ can be a quark or antiquark. 
Both next-to-leading order parton level cross sections contain 
divergences associated with collinear singularities in the initial state. 
These divergences can be made explicit in 
dimensional regularization with $d=4-2\epsilon$, they are process
independent and can be factorized into the bare parton distribution 
functions. In the case of the Drell--Yan process at ${\cal O}(\alpha_s)$,
all infinities can be absorbed into the bare quark distribution.

The use to dimensional regularization to compute spin dependent 
quantities faces the problem of representing $\gamma_5$ in 
$n\neq 4$ dimensions. We use the 
$\gamma_5$--prescription~\cite{HVBM} of 't~Hooft, Veltman, Breitenlohner 
and Maison (HVBM), which has been consistently 
used in the derivation of the next-to-leading
order corrections to the polarized splitting functions~\cite{nlosplit}.
In this prescription, one restricts the anticommutation property    
of $\gamma_5$ to the physical four dimensions, while $\gamma_5$ commutes in
the remaining $n-4$ dimensions. The major drawback of this formalism in the 
$\overline{{\rm MS}}$--scheme is the non--conservation of the 
flavour non--singlet axial vector currents~\cite{laring5} 
due to a non--vanishing first 
moment of the corresponding non--singlet NLO splitting function $\Delta
P_{qq,+}$. To restore the conservation of this non--singlet axial vector
current, a further scheme transformation~\cite{laring5}
 of the results obtained in the
HVBM formalism is needed~\cite{nlosplit}. An explicit formula for this 
scheme transformation, represented by a finite counterterm in the 
bare polarized quark distribution is given for example in~\cite{tgdy}. 

Inclusive 
vector boson production in unpolarized 
hadron collisions is up to ${\cal O}(\alpha_s)$
in perturbation theory identical to the Drell--Yan process, differences 
due to the coupling of the $Z^0$ boson to a closed quark loop~\cite{dym2}
occur only at ${\cal O}(\alpha_s^2)$. It should therefore be expected 
that the ${\cal O}(\alpha_s)$ corrections to the double polarized cross section
$\d \Delta \sigma_{LL} (y)/\d y$ are identical to the corrections 
to the $y$ distribution of lepton pairs produced  
in the polarized Drell--Yan process~\cite{weber,tgdy}. This identity 
could only be spoilt if the vector and axial-vector components 
of the vector boson couplings ($v_q$ and $a_q$) to 
polarized quarks would acquire different higher order corrections 
due to the breaking of the  anticommutation property of $\gamma_5$ 
in $n\neq 4$ dimensions.

Using the HVBM representation of 
$\gamma_5$, one can easily show that it is still possible 
to factor out the combination $v_q^2+a_q^2$ from the squared 
next-to-leading order matrix elements.
The resulting matrix elements then coincide with the 
matrix elements appearing in the polarized Drell--Yan process up to 
terms which yield only contributions of ${\cal O}(\epsilon)$
to the final result. The ${\cal O}(\alpha_s)$
corrections to the double polarized cross section~(\ref{eq:sigll}) are 
therefore indeed identical to the corrections to the polarized Drell--Yan 
process, as na\"{\i}vely expected.

The single polarized cross section defined 
in (\ref{eq:sigl}) vanishes due to parity conservation in the 'classical' 
Drell--Yan process, which is mediated by a virtual photon. Only the 
parity violating couplings of the $W^{\pm}$ and $Z^0$ bosons 
to quarks induce non-zero single spin asymmetries. In the lowest order 
annihilation process, one has to distinguish between the annihilation of 
a quark from the polarized hadron with an antiquark from the unpolarized 
hadron ($\Delta q+ \bar q$) and the annihilation of an antiquark from the 
polarized hadron with a quark from the unpolarized hadron ($\Delta \bar q +q$).
The $\epsilon$-dependent parts of these lowest order cross sections, 
relevant for the finite terms generated in the mass factorization, 
are non-identical if the HVBM prescription for $\gamma_5$ is used:
\begin{displaymath}
\Delta \hat{\sigma}_L (\Delta q + \bar q \to V) \sim -(1-\epsilon)\; , \qquad
\Delta \hat{\sigma}_L (\Delta \bar q + q \to V) \sim (1+\epsilon)\; .
\end{displaymath}

These normalization factors of the lowest order cross sections 
determine the terms that have to be factored out from all 
matrix elements of the 
${\cal O}(\alpha_s)$ subprocess contributions. The factor $-(1-\epsilon)$ is 
required for the ($\Delta q + \bar q$) annihilation process and for
the ($\Delta G + \bar q$) and $(\Delta q + G)$ gluon Compton scattering 
processes, while $(1+\epsilon)$ must be factored out from the 
($\Delta \bar q +  q$) annihilation process and the  ($\Delta G + q$)
and $(\Delta \bar q + G)$ gluon Compton scattering processes. Once the 
correct normalization has been factored out, the derivation of the 
partonic coefficient functions is identical to the derivation in 
the double polarized Drell--Yan process~\cite{tgdy}. We shall 
only quote the final results below.

The normalization factor common to all polarized and unpolarized 
cross sections is 
\begin{equation}
N = \frac{\pi G_F M_V^2\sqrt{2}}{3 S}\; , 
\end{equation}
and the coupling factors read
\begin{eqnarray}
& c_{ij} = |V_{ij}| & \mbox{for $W^{\pm}$}\; ,\nonumber \\
& c_{ij} = (v_i^2 + a_i^2) \delta_{ij} & \mbox{for $Z^0$ unpolarized and 
double polarized} \; ,\nonumber \\
& c_{ij} = 2v_i a_i \delta_{ij} & \mbox{for $Z^0$  
single polarized}\; .
\end{eqnarray}
The unpolarized and polarized vector boson production 
cross sections at next-to-leading order 
can be expressed in a compact analytic form as convolution of parton 
distributions with partonic coefficient functions:
\begin{eqnarray}
\frac{\d \sigma}{\d y} & = &  N \sum_{i,j} c_{ij} 
\int_{x_1^0}^1 \d x_1 \int_{x_2^0}^1 \d x_2 \nonumber \\
& &  \hspace{-1.6cm}
\times \Bigg\{\left[ 
D_{q\bar q}^{(0)} (x_1,x_2,\xaa,\xbb)  +\frac{\alpha_s}{2\pi}
D_{q\bar{q}}^{(1)}
\left(x_1,x_2,\xaa,\xbb,\frac{M^2}{\mu_F^2}\right)\right]\nonumber \\
& & \hspace{-0.4cm} \times
\Big\{ q_i(x_1,\mu_F^2) \bar{q}_j(x_2,\mu_F^2) +  
\bar{q}_i(x_1,\mu_F^2) q_j(x_2,\mu_F^2) \Big\} \nonumber \\
& & \hspace{-1.2cm} +  \frac{\alpha_s}{2\pi}
D_{gq}^{(1)} \left(x_1,x_2,\xaa,\xbb, 
\frac{M^2}{\mu_F^2}\right) G(x_1,\mu_F^2) \left\{ 
q_j(x_2,\mu_F^2) + 
\bar{q}_j (x_2,\mu_F^2) \right\} \nonumber \\
& & \hspace{-1.2cm} + \frac{\alpha_s}{2\pi}
D_{qg}^{(1)} \left(x_1,x_2,\xaa,\xbb, 
\frac{M^2}{\mu_F^2}\right) \left\{ 
q_i(x_1,\mu_F^2) + 
\bar{q}_i (x_1,\mu_F^2) \right\} G(x_2,\mu_F^2) \Bigg\}\; , \\
\frac{\d \Delta \sigma_{LL}}{\d y} & = & - N \sum_{i,j} c_{ij} 
\int_{x_1^0}^1 \d x_1 \int_{x_2^0}^1 \d x_2 \nonumber \\
& &  \hspace{-1.6cm}
\times \Bigg\{\left[ 
D_{q\bar q}^{(0)} (x_1,x_2,\xaa,\xbb)  +\frac{\alpha_s}{2\pi}
D_{q\bar{q}}^{(1)}
\left(x_1,x_2,\xaa,\xbb,\frac{M^2}{\mu_F^2}\right)\right]\nonumber \\
& & \hspace{-0.4cm} \times
\Big\{ \Delta q_i(x_1,\mu_F^2) \Delta \bar{q}_j(x_2,\mu_F^2)   +
\Delta \bar{q}_i(x_1,\mu_F^2) \Delta q_j(x_2,\mu_F^2) \Big\} \nonumber \\
& & \hspace{-1.2cm} + \frac{\alpha_s}{2\pi}
\Delta D_{gq}^{(1)} \left(x_1,x_2,\xaa,\xbb, 
\frac{M^2}{\mu_F^2}\right) \Delta G(x_1,\mu_F^2) \left\{ 
\Delta q_j(x_2,\mu_F^2) + 
\Delta \bar{q}_j (x_2,\mu_F^2) \right\} \nonumber \\
& & \hspace{-1.2cm} +  \frac{\alpha_s}{2\pi}
\Delta D_{qg}^{(1)} \left(x_1,x_2,\xaa,\xbb, 
\frac{M^2}{\mu_F^2}\right) \left\{ 
\Delta q_i(x_1,\mu_F^2) + 
\Delta \bar{q}_i (x_1,\mu_F^2) \right\} \Delta 
G(x_2,\mu_F^2)  \Bigg\}\! , \label{eq:llmas}\\
\frac{\d \Delta \sigma_{L}}{\d y} & = & N \sum_{i,j} c_{ij} 
\int_{x_1^0}^1 \d x_1 \int_{x_2^0}^1 \d x_2 \nonumber \\
& &  \hspace{-1.6cm}
\times \Bigg\{\left[ 
D_{q\bar q}^{(0)} (x_1,x_2,\xaa,\xbb)  +\frac{\alpha_s}{2\pi}
D_{q\bar{q}}^{(1)}
\left(x_1,x_2,\xaa,\xbb,\frac{M^2}{\mu_F^2}\right)\right]\nonumber \\
& & \hspace{-0.4cm} \times
\Big\{ - \Delta q_i(x_1,\mu_F^2) \bar{q}_j(x_2,\mu_F^2) +  
\Delta \bar{q}_i(x_1,\mu_F^2) q_j(x_2,\mu_F^2) \Big\} \nonumber \\
& & \hspace{-1.2cm} + \frac{\alpha_s}{2\pi}
\Delta D_{gq}^{(1)} \left(x_1,x_2,\xaa,\xbb, 
\frac{M^2}{\mu_F^2}\right) \Delta G(x_1,\mu_F^2) \left\{ 
q_j(x_2,\mu_F^2) - 
\bar{q}_j (x_2,\mu_F^2) \right\} \nonumber \\
& & \hspace{-1.2cm} +  \frac{\alpha_s}{2\pi}
D_{qg}^{(1)} \left(x_1,x_2,\xaa,\xbb, 
\frac{M^2}{\mu_F^2}\right) \left\{ 
- \Delta q_i(x_1,\mu_F^2) + 
\Delta \bar{q}_i (x_1,\mu_F^2) \right\}  
G(x_2,\mu_F^2)  \Bigg\}\; .\label{eq:lmas}
\end{eqnarray}
The partonic coefficient functions in the above expressions are identical to
the coefficient functions occurring in the $y$-distributions of the 
polarized and unpolarized Drell--Yan 
process~\cite{kubar,weber,tgdy,smrs}. They read:
\begin{eqnarray}
D_{q\bar q}^{(0)} (x_1,x_2,\xaa,\xbb) & = & \delta (x_1-x_1^0)\, 
\delta (x_2-x_2^0) \; , \\
D_{q\bar{q}}^{(1)}
\left(x_1,x_2,\xaa,\xbb,\frac{M^2}{\mu_F^2}\right)
& = & C_F \Bigg\{ \delta (x_1-x_1^0) \,
\delta (x_2-x_2^0) \bigg[ \frac{\pi^2}{3} - 8 + 2 \Li_2 (\xaa ) 
+ 2 \Li_2 ( \xbb ) 
\nonumber \\ 
& & \hspace{0.6cm}
+\ln^2 (1-\xaa) + \ln^2 (1-\xbb) + 2 \ln \frac{\xaa}{1-\xaa} \ln 
\frac{\xbb}{1-\xbb} \bigg]  \nonumber \\
& & + \Bigg( \delta (\xa-\xaa) \bigg[ \frac{1}{\xb} 
- \frac{\xbb}{\xb^2} - \frac{\xbb\,^2+\xb^2}{\xb^2(\xb-\xbb)} \ln 
\frac{\xbb}{\xb} \nonumber \\
& & \hspace{0.6cm}
+ \frac{\xbb\,^2+\xb^2}{\xb^2} 
\left(\frac{\ln (1-\xbb/\xb)}{\xb-\xbb}\right)_{+}
+ \frac{\xbb\,^2+\xb^2}{\xb^2} \frac{1}{\left(\xb-\xbb\right)_{+}}
\nonumber\\ && \hspace{0.6cm} \ln \frac
{2\xbb(1-\xaa)}{\xaa(\xb+\xbb)} 
\bigg] + (1 \leftrightarrow 2 ) \Bigg) \nonumber \\
& & + \frac{
G^A(\xa,\xb,\xaa,\xbb)}{\left[(\xa-\xaa)(\xb-\xbb)\right]_{+}} + 
H^A (\xa,\xb,\xaa,\xbb) 
\nonumber \\
&& 
+ \ln \frac{M^2}{\mu_F^2} \Bigg\{  \delta (x_1-x_1^0)\, 
\delta (x_2-x_2^0) \bigg[ 3 + 2 \ln \frac{1-\xaa}{\xaa} + 2 \ln 
\frac{1-\xbb}{\xbb}
\bigg] \nonumber\\
&& \hspace{0.6cm}
+ \bigg( \delta (\xa-\xaa)  
 \frac{\xbb\,^2+\xb^2}{\xb^2}\frac{1}{\left(\xb-\xbb\right)_{+}} + (1 
\leftrightarrow 2 ) \bigg) \Bigg\} \Bigg\}\; ,   \\
D_{gq}^{(1)}
\left(x_1,x_2,\xaa,\xbb,\frac{M^2}{\mu_F^2}\right) & = & 
T_F\Bigg\{\frac{\delta (\xb-\xbb) }{\xa^3} \Bigg[ (\xaa\,^2 +(\xa-\xaa)^2) 
\ln\frac{2(\xa-\xaa)(1-\xbb)}{(\xa+\xaa)\xbb} 
 \nonumber \\
& & \hspace{0.6cm}
+ 2\xaa(\xa-\xaa)\Bigg] 
+ \frac{G^C(\xa,\xb,\xaa,\xbb)}{(\xb-\xbb)_{+}} + 
H^C(\xa,\xb,\xaa,\xbb) 
\nonumber \\ & & 
+  \ln \frac{M^2}{\mu_F^2} \Bigg\{ \frac{\delta (\xb-\xbb) }{\xa^3}
 (\xaa\,^2 +(\xa-\xaa)^2) \Bigg\}\Bigg\}\; , \\
D_{qg}^{(1)}
\left(x_1,x_2,\xaa,\xbb,\frac{M^2}{\mu_F^2}\right) & = & 
D_{gq}^{(1)}
\left(x_2,x_1,\xbb,\xaa,\frac{M^2}{\mu_F^2}\right) \; , \\
\Delta D_{gq}^{(1)}
\left(x_1,x_2,\xaa,\xbb,\frac{M^2}{\mu_F^2}\right) & = & 
T_F\Bigg\{\frac{\delta (\xb-\xbb) }{\xa^2} \Bigg[ (2\xaa -\xa) 
\ln\frac{2(\xa-\xaa)(1-\xbb)}{(\xa+\xaa)\xbb} 
 \nonumber \\
& & \hspace{0.6cm}
+ 2(\xa-\xaa)\Bigg] 
+ \frac{\Delta G^C(\xa,\xb,\xaa,\xbb)}{(\xb-\xbb)_{+}} + 
H^C(\xa,\xb,\xaa,\xbb) 
\nonumber \\ & & 
+  \ln \frac{M^2}{\mu_F^2} \Bigg\{ \frac{\delta (\xb-\xbb) }{\xa^2}
 (2\xaa -\xa) \Bigg\}\Bigg\}\; , \\
\Delta D_{qg}^{(1)}
\left(x_1,x_2,\xaa,\xbb,\frac{M^2}{\mu_F^2}\right) & = & 
\Delta D_{gq}^{(1)}\left(x_2,x_1,\xbb,\xaa,\frac{M^2}{\mu_F^2}\right)\; ,
\end{eqnarray}
where
\begin{eqnarray}
G^A(\xa,\xb,\xaa,\xbb) & = & \frac{2 (\xa\xb+\xaa\xbb)
(\xaa\,^2\xbb\,^2+\xa^2\xb^2)}
{\xa^2\xb^2(\xa+\xaa)(\xb+\xbb)}\; , \nonumber \\
H^A (\xa,\xb,\xaa,\xbb) & = & 
-\frac{4\xaa\xbb(\xaa\xbb+\xa\xb)}{\xa\xb(\xa\xbb+\xb\xaa)^2}\; , \nonumber \\
G^C(\xa,\xb,\xaa,\xbb) & = & \frac{2\xbb(\xaa\,^2\xbb\,^2+(\xaa\xbb-\xa\xb)^2)
(\xaa\xbb+\xa\xb)}{\xa^3\xb^2(\xa\xbb+\xb\xaa)(\xb+\xbb)}\; ,
\nonumber \\
H^C (\xa,\xb,\xaa,\xbb) & = &\frac{2\xaa\xbb(\xaa\xbb+\xa\xb)(\xa\xaa\xb^2+
\xaa\xbb(\xa\xbb+2\xaa\xb))}{\xa^2\xb^2(\xa\xbb+\xb\xaa)^3}\; , \nonumber\\  
\Delta G^C(\xa,\xb,\xaa,\xbb) 
& = & \frac{2\xbb(2\xaa\xbb-\xa\xb)(\xaa\xbb+\xa\xb)}
{\xa^2\xb(\xa\xbb+\xb\xaa)(\xb+\xbb)}\; .
\nonumber 
\end{eqnarray}  

The occurrence of double polarized and unpolarized coefficient functions 
in the single polarized vector boson production
cross section can be understood by simple arguments. The 
$W$-boson couples only 
to one particular helicity of the incoming quarks. The unpolarized,
single and double polarized $q\bar q$
cross sections receive therefore only 
a single contribution from one particular configuration 
of quark and antiquark helicities and are thus identical up to an overall 
sign. In the case of 
quark-gluon Compton scattering, both different gluon helicities 
can contribute; the $qg$ cross sections thus receive contributions 
from two different helicity combinations and the unpolarized and 
double polarized parton cross sections are different. In the case 
of the single polarized cross section, it is easy to identify the 
polarized quark-unpolarized gluon contribution with the  
fully unpolarized subprocess, and the polarized gluon-unpolarized 
quark contribution with the fully polarized subprocess. These 
identifications are less obvious for the $Z^0$ boson production, but can 
be made explicit if the $Z^0$ couplings to the different quark 
handednesses are separated.

The partonic coefficient functions appearing in the single 
polarized vector boson cross sections have been first studied by Weber 
in~\cite{weber2}, where they were obtained as a by-product in the 
calculation of the soft gluon resummation to these single polarized 
cross sections. Our results are in a simpler form than the ones given 
in~\cite{weber2} and differ both in the polarized gluon-unpolarized quark and 
the polarized quark-unpolarized gluon subprocesses. The discrepancy
in the former coefficient can be attributed to a different factorization 
scheme used for the polarized gluon distribution in~\cite{weber2}; the 
discrepancy in the latter can only be understood to be due to a 
non-conventional normalization of the number of unpolarized gluon 
states in~\cite{weber2}. A consistency check of our results is 
given by the quark helicity arguments discussed above.  These  
explain the occurrence of the unpolarized and double polarized coefficient 
functions in the expression for the double polarized cross section.
Moreover, integration of our results over $y$ yields the QCD corrections 
to the inclusive asymmetries, which have been recently calculated 
by Kamal 
in~\cite{kamal2}. Expressing the final results of~\cite{kamal2} in the 
$\overline{{\rm MS}}_{HC}$-scheme (which is identical to our implementation 
of the $\overline{{\rm MS}}$ scheme), we find complete agreement with 
our results integrated over $y$. 

\section{Numerical results}
\label{sec:num}
The numerical impact of the next-to-leading order corrections 
derived in the previous section can be illustrated by evaluating the 
asymmetries (\ref{eq:defll}) and (\ref{eq:defl}) using recent next-to-leading 
order parameterizations of polarized and  unpolarized parton distributions. 

When using the polarized GS parton distribution 
functions~\cite{gs},
we take $\Lambda^{{\rm QCD}}_{n_f=4}=231$~MeV, the 
corresponding unpolarized cross sections 
are then evaluated using the unpolarized MRS parton
distribution functions set A$'$~\cite{mrsap}. The polarized 
GRSV distributions~\cite{grsv} are 
consistently used in combination with the unpolarized distributions 
from GRV~\cite{grv} and for
$\Lambda^{{\rm QCD}}_{n_f=4}=200$~MeV. If not stated otherwise, we shall 
always use $\mu_F=M$; the strong coupling constant $\alpha_s$ is evaluated 
at $\mu_F$. All results in this section are obtained for 
longitudinally polarized proton-proton collisions at RHIC ($\sqrt S=500$~GeV);
the single spin asymmetries are always obtained for the configuration where 
the polarized proton is moving in the $+y$ direction. 
\begin{figure}[t]
\begin{center}
~ \epsfig{file=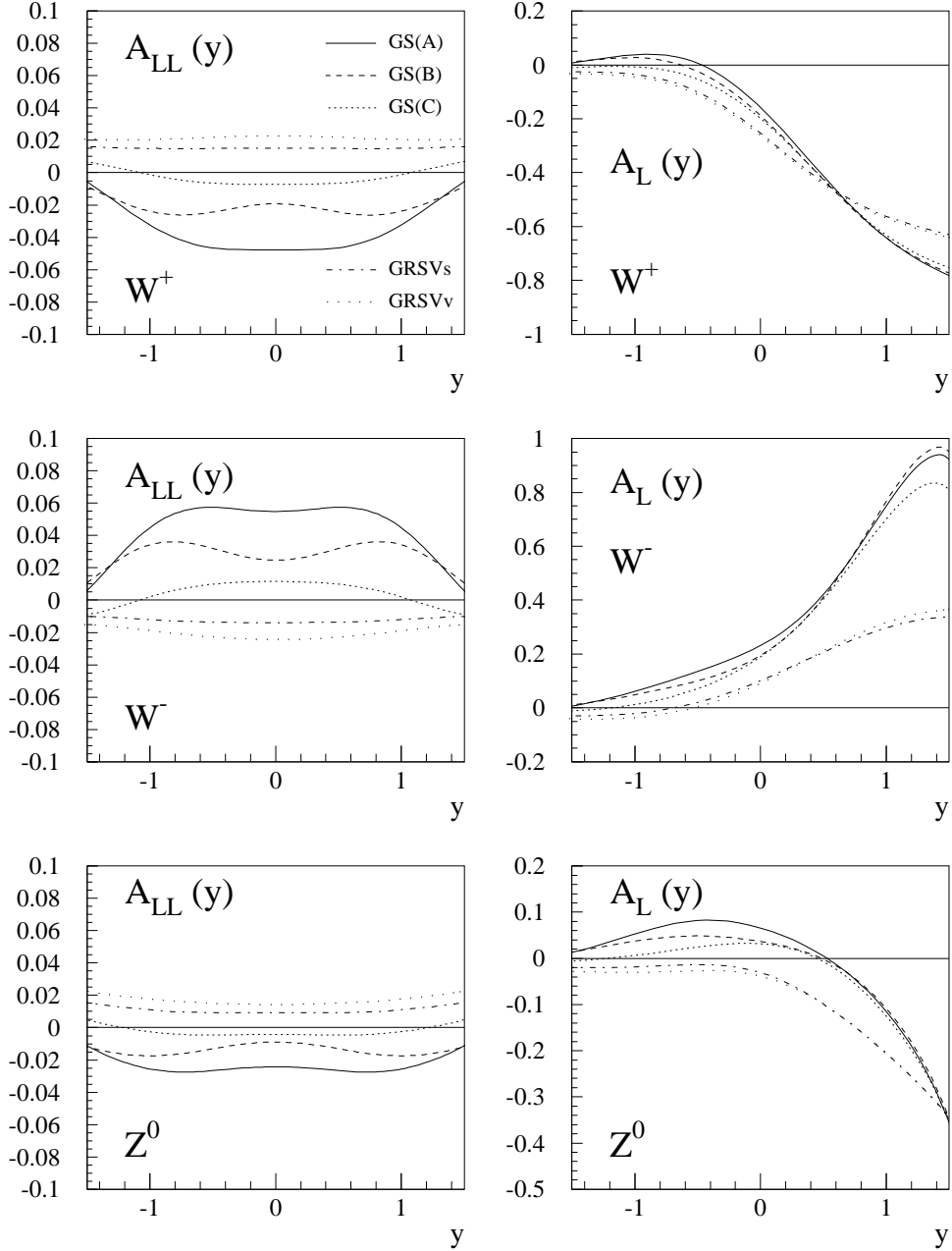,width=13cm}
\caption{
Predictions for double (left) and single (right) spin asymmetries in 
vector boson production in 
$pp$ collisions at $\sqrt s = 500$~GeV obtained for different parameterizations
of the next-to-leading order polarized parton densities.
}
\label{fig:pdf}
\end{center}
\end{figure}

It must be kept in mind that all present day parameterizations 
of the polarized parton distributions are fitted only to data on the 
structure function $g^{p,d,n}_1(x,Q^2)$. These data only provide very loose 
constraints on the polarized sea quark distributions at large $x$ and 
are clearly insufficient for a flavour decomposition of the polarized light 
quark sea. Consequently, our present knowledge on the polarized sea 
quark distributions is very poor. This uncertainty is illustrated in 
Figure~\ref{fig:pdf} showing the vector boson production 
asymmetries $A_{LL}(y)$ and $A_L(y)$ 
evaluated at next-to-leading order
using the polarized parton distribution functions of 
GS(A--C)~\cite{gs} and GRSVs,v~\cite{grsv}. A more extensive study at
lowest order has recently been presented in~\cite{soffernew}.  

Based on these present parameterizations, it is not possible to predict 
sign or magnitude of the double spin asymmetry $A_{LL}(y)$ or of the 
single spin asymmetry $A_L(y<0)$; both asymmetries being dominated 
by contributions from the polarized sea distributions in the nucleon.
From the ratio of polarized to unpolarized sea quark distributions 
in present parameterizations, it can only be estimated that 
$|A_{LL}(y)| \lapprox 0.06$ and $|A_{L}(y<0)| \lapprox 0.3$. 
Only $A_L(y>0)$ is dominated by the behaviour of the polarized valence 
quarks at relatively large $x$, which is reasonably
well constrained by the present $g^{p,d,n}_1(x,Q^2)$ data. 
Consequently, the different 
parameterizations yield similar predictions for $A_L(y>0)$, a sizable 
discrepancy can only be seen for $W^-$ production. This discrepancy is 
related to a still sizable difference in the $\Delta d_v(x,Q^2)$ distributions 
in the GRSV and GS parametrizations and will most likely 
be sorted out in the near future when new precision data on the 
deuteron spin structure function $g_1^d(x,Q^2)$ will become available. 

The above discussion clearly shows that our present knowledge 
on the polarized quark distributions is still
insufficient for  precise 
predictions of asymmetries in vector boson production at RHIC. 
The 
following studies of the numerical effects of the 
next-to-leading order corrections can therefore only illustrate the
qualitative effects of these corrections without making definite,
quantitative
predictions for asymmetries or $K$-factors.

\begin{figure}[t]
\begin{center}
~ \epsfig{file=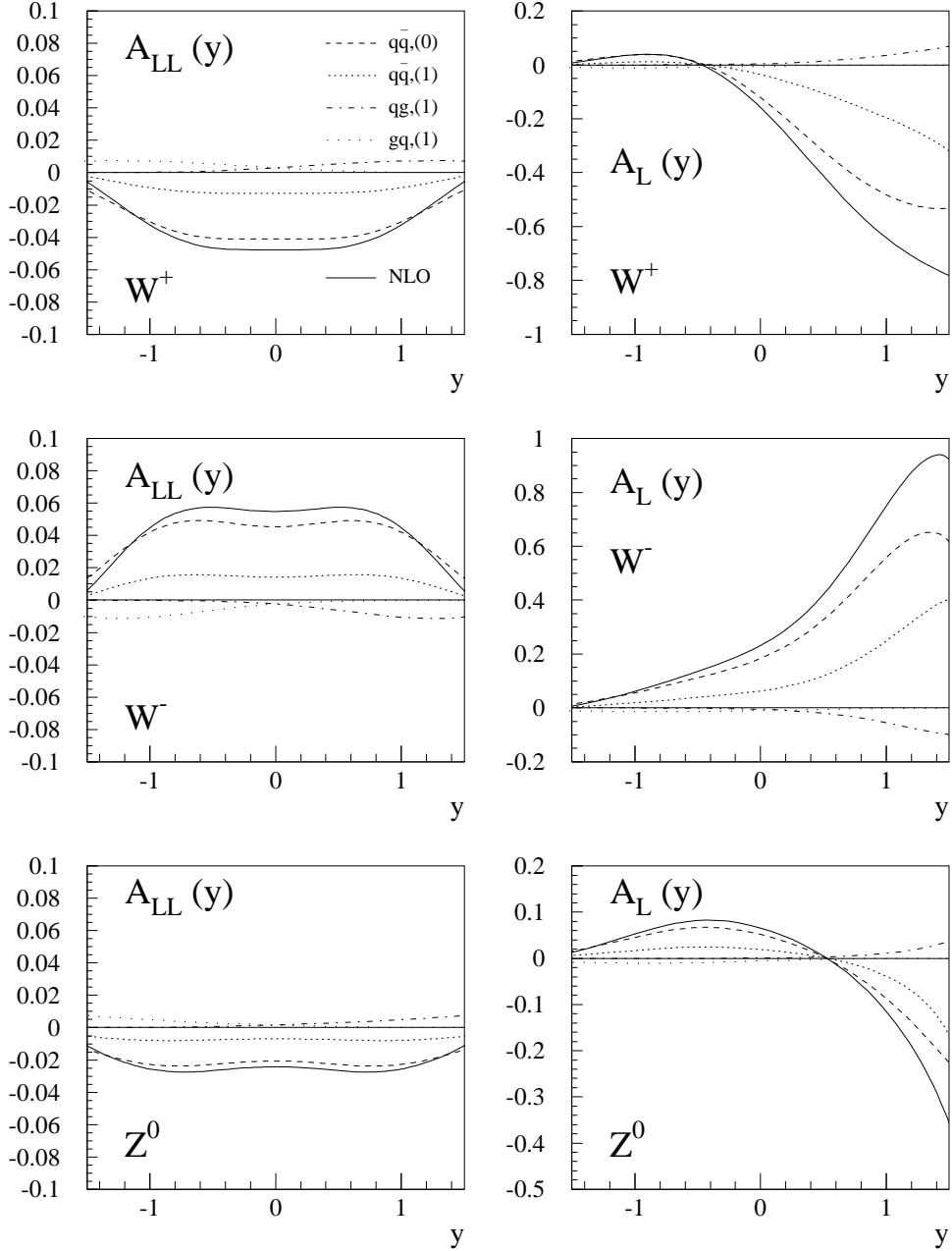,width=13cm}
\caption{
Contributions of the individual parton level subprocesses to the 
double (left) and single (right) polarized vector boson production cross 
sections in $pp$ collisions at $\sqrt s = 500$~GeV. All polarized 
cross sections are normalized to the full unpolarized cross section at 
next-to-leading order.
}
\label{fig:nlo}
\end{center}
\end{figure}
The contributions of the individual subprocesses ($q\bar q$--annihilation
at leading and next-to-leading order and quark-gluon Compton scattering) 
to the polarized vector boson cross sections are shown in 
Fig.~\ref{fig:nlo}. All polarized subprocess 
cross sections are obtained with the polarized GS(A) 
parton distributions and are normalized to the full unpolarized 
cross section at next-to-leading order. In the case of the quark-gluon 
Compton process, we distinguish moreover the case where the gluon 
originates from the proton moving in $-y$ direction ($qg$) and the 
case where the gluon originates from the proton moving in $+y$ direction
($gq$). This distinction is irrelevant for $\Delta \sigma_{LL}(y)$, 
it becomes however important for $\Delta \sigma_{L}(y)$, 
where the $qg$ process 
probes $\Delta q(\xa,Q^2)\, G(\xb,Q^2)$  while 
the $gq$ process is sensitive on $\Delta G(\xa,Q^2)\, q(\xb,Q^2)$.

\begin{figure}[t]
\begin{center}
~ \epsfig{file=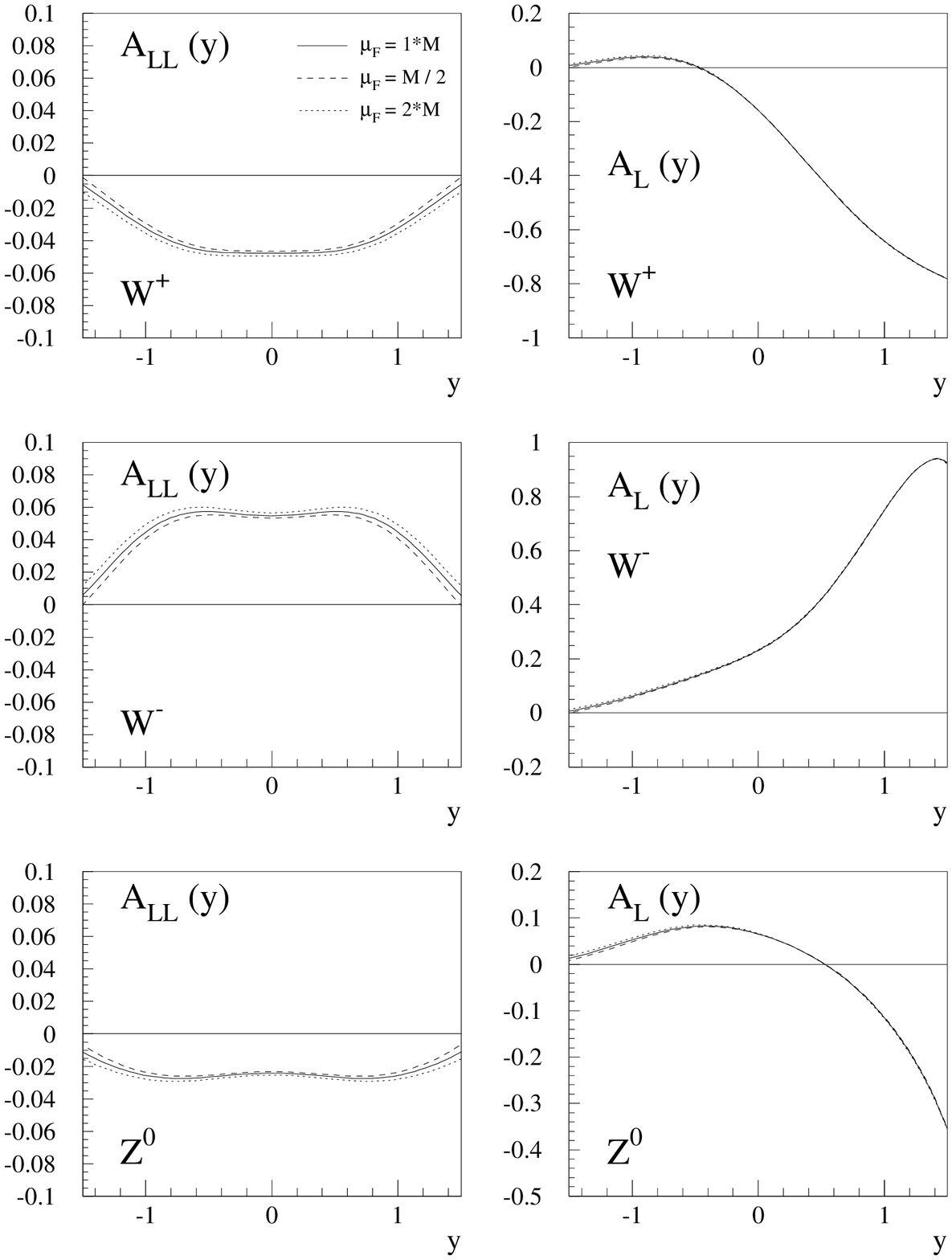,width=13cm}
\caption{
Variation of the double (left) and single 
(right) spin asymmetries in vector boson production in 
$pp$ collisions at $\sqrt s = 500$~GeV under variation of the mass
factorization scale.
}
\label{fig:muf}
\end{center}
\end{figure}
The numerical impact of the individual contributions to both 
$\Delta \sigma_{L}(y)$ and $\Delta \sigma_{LL}(y)$ 
is similar to the impact of these contributions in the ordinary 
polarized Drell--Yan process~\cite{tgdy}.  
The ${\cal O}(\alpha_s)$ correction to the $q\barq$--annihilation process 
enhances significantly the lowest order prediction while the quark--gluon 
Compton process contributes with a sign opposite to the annihilation process. 
A definite
prediction of the relative magnitude of these two next-to-leading 
order processes is however only possible for $\Delta \sigma_{L}(y>0)$,
as we shall explain below.

The quark-gluon Compton contributions to  $\Delta \sigma_{L}(y)$ arise 
mainly from the $qg$ 
configuration for $y>0$ and from the $gq$ configuration for 
$y<0$. Consequently, an extraction of the large $x$ behaviour of the 
polarized quark distributions at next-to-leading order
from a measurement of $A_L(y>0)$ will not 
suffer from the uncertainty on the polarized gluon distribution, since only
the known unpolarized gluon distribution contributes in the $qg$ 
configuration. The numerical 
magnitude of the next-to-leading order corrections 
to $A_L(y>0)$ can therefore predicted reliably: inclusion of these corrections
enhances $|\Delta \sigma_L(y>0)|$ by about 15--25\% compared to its lowest 
order value.  

A quantification of the next-to-leading 
corrections to  $\Delta \sigma_{LL}(y)$ and  $\Delta \sigma_{L}(y<0)$
requires on the other hand some knowledge on the magnitude of the 
polarized gluon distribution $\Delta G(x,Q^2)$, which enters these 
cross sections in the quark-gluon Compton scattering process. The same 
problem was already encountered in the QCD corrections to the 
longitudinally polarized Drell--Yan process~\cite{tgdy}. 
It illustrates that any
extraction of the polarized sea quark distributions at next-to-leading 
order requires at least some order-of-magnitude knowledge on the 
polarized gluon distribution.  

The change of physical 
quantities such as cross sections or asymmetries under variations of the 
(unphysical) mass factorization scale $\mu_F$ enables an estimate of the 
numerical importance of unknown higher order corrections. We quantify
this uncertainty in Fig.~\ref{fig:muf}, showing the double and single 
spin asymmetries in vector boson production
for $\mu_F=0.5,\,1,\,2\, M_V$. 
It can be seen that
the absolute value of $A_{LL}(y)$ changes by less
than 0.005 in the central region, the variations become slightly 
larger in the region where $y$ approaches its kinematic limit, but they 
never exceed 0.01. The variations in $A_L(y>0)$ are even smaller and 
do not exceed 0.004, they become larger in $A_L(y<0)$,
where they can amount up to 0.01. All these variations are however 
significantly smaller than the difference between predictions 
for $A_{LL}(y)$ and $A_L(y)$ obtained with various parameterizations 
of the polarized parton distribution functions (cf.~Fig.~\ref{fig:pdf}). 
It must be emphasized that the relative smallness  of these  
observed variations with the mass factorization scale 
arises partly due to the fact that the scale dependence 
of polarized and unpolarized cross sections largely compensate when taking 
their ratio in the asymmetry. The unpolarized 
vector boson production cross sections vary by about 15\% between
 $\mu_F=0.5\, M_V$ and $\mu_F=2\, M_V$.

\section{Conclusions}
\label{sec:conc}
We have presented a complete calculation of the 
perturbative ${\cal O}(\alpha_s)$ 
corrections to double ($A_{LL}(y)$)
and single ($A_L(y)$) spin 
asymmetries in massive
vector boson production
in collisions of longitudinally polarized hadrons. 
The results can be expressed in simple analytic form and are given 
in eqs.(\ref{eq:llmas}) and (\ref{eq:lmas}).
These corrections are of particular importance for 
the extraction of the polarized sea quark distributions from measurements 
of vector boson production asymmetries at RHIC. 

The next-to-leading order terms contributing 
to $A_{LL}(y)$ agree with the corrections to this asymmetry in 
the longitudinally polarized Drell--Yan process~\cite{weber,tgdy}, as naively 
expected. 

Corrections to $A_L(y)$
had been first considered by Weber~\cite{weber2} in the context of a 
soft gluon resummation to single spin asymmetries.
While our results for the $\Delta q+\bar q$, $\Delta \bar q +q$ and 
$\Delta G + q$ 
next-to-leading order 
subprocesses agree (after suitable change of factorization scheme)
with~\cite{weber2}, we obtain a different result for the 
$\Delta q + G$ subprocess (polarized quark--unpolarized gluon). This 
difference could be attributed to a non-conventional normalization of 
the gluonic polarization sum used in~\cite{weber2}. A check of 
our results is given by the fact that all 
single polarized subprocess cross sections 
can be directly
related to subprocess cross sections in the double polarized 
or unpolarized case. This behaviour should be expected due to the 
helicity structure of the vector boson couplings to quarks.
A further check is the 
integration of our results over the rapidity $y$, which 
reproduces the recent results of~\cite{kamal2} while showing the 
same above 
discrepancy with~\cite{weber2}.

We have demonstrated that the 
numerical impact of these corrections on the polarized 
vector boson cross sections is very similar to the impact of the 
next-to-leading order corrections in the unpolarized case.
The 
corrections due to the individual subprocesses
turn out to be sizable. However, a $K$--factor
between leading and next-to-leading order results can at present
only be predicted for $A_L(y>0)$, which is insensitive to the yet 
unknown magnitude of the polarized gluon distribution.

With the knowledge of the next-to-leading order corrections, it is 
furthermore possible 
to quantify the uncertainty of the theoretical prediction due to 
the choice of factorization scale
and hence the 
theoretical error on a measurement of the polarized quark distributions. 
We have demonstrated that the absolute value of $A_{LL}(y)$ varies by 
less than 0.005 in the central region under change of the 
factorization scale. The variations in $A_L(y)$ are only sizable for $y<0$,
where they amount up to 0.01. These variations are however significantly 
smaller than the difference between predictions obtained for 
$A_{LL}(y)$ and $A_L(y)$ using different parameterizations for the
polarized parton distribution functions.

To summarize, the calculations presented in this paper allowed us to 
investigate two possible sources of theoretical uncertainty on 
an extraction of the polarized 
quark distributions from double and single spin 
asymmetries in vector boson 
production at RHIC. 
 We found that the variation of the mass  
factorization scale had only minor effects on all asymmetries, thus 
indicating that the effect from unknown higher order terms can be 
expected to be small. The effects on these asymmetries from the 
yet unknown polarized gluon distribution were on the other hand found 
to be sizable for both $A_{LL}(y)$ and $A_L(y<0)$, while being negligible 
in $A_L(y>0)$. This observation illustrates that a
consistent extraction of the
next-to-leading order polarized sea quark distributions from vector 
boson production at RHIC will require at least some order-of-magnitude 
information on the polarized gluon distribution, which is to be obtained
from other processes.

\section*{Acknowledgements}
\noindent
Part of the work presented in this paper has been carried out 
during a visit to the Fermilab Theory Group, whose kind hospitality is 
gratefully acknowledged. The author would moreover like to thank 
W.~Vogelsang for clarifying discussions on Ref.~\cite{weber2}.
\goodbreak

\end{document}